\documentclass[reprint,amsmath,amssymb,aps]{revtex4-2}

\newcommand{\RomanNumeralCaps}[1]
    {\MakeUppercase{\romannumeral #1}}

\usepackage{graphicx}
\usepackage{hyperref}
\usepackage{dcolumn}
\usepackage{bm}
\usepackage{color}
\usepackage{textcomp, gensymb}
\usepackage{amssymb}
\DeclareUnicodeCharacter{2212}{-}

\begin{document}

\preprint{APS/123-QED}

\title{Strongly correlated Hofstadter subbands in minimally twisted bilayer graphene}

\author{Cheng Shen$^{1,a\dagger}$}
\author{Yifei Guan$^{1\dagger}$}
\author{Davide Pizzirani$^{2}$}
\author{Zekang Zhou$^{1}$}
\author{Punam Barman$^{1}$}
\author{Kenji Watanabe$^{3}$}
\author{Takashi Taniguchi$^{4}$}
\author{Steffen Wiedmann$^{2}$}
\author{Oleg V. Yazyev$^{1}$}
\author{Mitali Banerjee$^{1\star}$}

\affiliation{$^{1}$Institute of Physics, Ecole Polytechnique Fédérale de Lausanne (EPFL), CH-1015 Lausanne, Switzerland}

\affiliation{$^{2}$High Field Magnet Laboratory (HFML-EMFL), Radboud University, Toernooiveld 7, 6525 ED Nijmegen, The Netherlands}

\affiliation{$^{3}$Research Center for Functional Materials, National Institute for Materials Science, 1-1 Namiki, Tsukuba 305-0044, Japan}

\affiliation{$^{4}$International Center for Materials Nanoarchitectonics, National Institute for Materials Science, 1-1 Namiki, Tsukuba 305-0044, Japan}

\affiliation{$^{a}$ Current address: School of Physics, University of Electronic Science and Technology of China, Chengdu 610054, China}
\affiliation{$^{\dagger}$ these authors contribute equally to the work}
\affiliation{$^{\star}$ corresponding author: mitali.banerjee@epfl.ch}  

\begin{abstract}
Moiré superlattice in twisted bilayer graphene has been proven to be a versatile platform for exploring exotic quantum phases. Extensive investigations have been invoked focusing on the zero-magnetic-field phase diagram at the magic twist angle around $\theta=1.1\degree$, which has been indicated to be an exclusive regime for exhibiting flat band with the interplay of strong electronic correlation and untrivial topology in the experiment so far. In contrast, electronic bands in non-magic-angle twisted bilayer graphene host dominant electronic kinetic energy compared to Coulomb interaction. By quenching the kinetic energy and enhancing Coulomb exchange interactions by means of an applied perpendicular magnetic field, here we unveil gapped flat Hofstadter subbands at large magnetic flux that yield correlated insulating states in minimally twisted bilayer graphene at $\theta=0.41\degree$. These states appear with isospin symmetry breaking due to strong Coulomb interactions. Our work provides a platform for studying the phase transition of the strongly correlated Hofstadter spectrum.
 
\end{abstract}

\maketitle

\section*{Introduction}

Electrons in solids undergo a phase transition from conducting to a localized state when on-site Coulomb interaction dominates over the kinetic hopping energy. Such a phase transition, typically described by the Mott-Hubbard model, requires a flat electronic band with a quenched kinetic energy. Recent progress in moiré superlattice has revealed twisted bilayer graphene (TBG) as a platform system, in which a flat electronic band can arise from periodic interlayer coupling, and a comparable Coulomb interaction energy is approachable by tuning moiré wavelength through the tunable twist angle \cite{Cao2018CorrelatedSuperlattices, Cao2018}. Diverse quantum states with strong electronic correlation have been revealed in this system including correlated insulator \cite{Cao2018CorrelatedSuperlattices}, superconductivity \cite{Cao2018}, orbital magnet \cite{Sharpe2019EmergentGraphene, Lu2019SuperconductorsGraphene, Serlin2020IntrinsicHeterostructure, Tschirhart2021ImagingInsulator} and fractional Chern insulator \cite{Xie2021FractionalGraphene}. Manipulating electronic flat bands in moiré superlattice provides the alluring potential to approach exotic quantum phases in condensed matters. To date, most work on TBG has focused on the so-called ‘magic angle’ of approximately $1.1\degree$, where flat bands are intrinsically emerging. With a fairly more dispersive band, the non-magic angle TBG typically requires external tuning to enable strong electronic correlation \cite{Yankowitz1059, Arora2020SuperconductivityWSe2}. 

Subjected to an out-of-plane magnetic field, the magnetic length $l_B$ provides the two-dimensional electron gas with tunable strength of Coulomb interactions. For electrons in a periodic lattice, the interplay between magnetic cyclotron energy and the periodic lattice potential gives rise to a non-interacting fractal Hofstadter butterfly spectrum with a recursive magnetic subband structure. The Coulomb interaction is intuitively expected to affect such Hofstadter butterfly structure, yet not fully conclusive \cite{Czajka2006HofstadterSystem, Barelli1996DoubleModel, Doh1998EffectsSpectrum, Ding2022ThermodynamicsField}. Adding an interacting Hubbard energy to Hofstadter Hamiltonian, the so-called Hofstadter-Hubbard model, is constructed to understand the correlated Hofstadter electrons.  Experimentally, previous investigations have exhibited a non-interacting fractal spectrum by magnetoresistance oscillations in the graphene/hexagonal boron nitride (hBN) moiré heterostructure \cite{Hunt2013MassiveHeterostructure, Ponomarenko2013CloningSuperlattices, Dean2013HofstadtersSuperlattices}, and now extended to a strongly correlated regime in twisted graphene superlattice.  For the latter case, the inherited strong Coulomb interaction from zero magnetic field results in the isospin-symmetry-broken Hofstadter subband ferromagnetism and translation-symmetry broken Chern insulator phases under small magnetic flux \cite{Saito2021HofstadterGraphene, Yu2022CorrelatedGraphene}. Albeit the Hofstadter butterfly picture is generic to all twist angles, studies of correlated Hofstadter subbands in small twist angle TBG samples and the impact of Coulomb interaction associated with magnetic field are scarce. Here, we show the phase transition of Hofstadter subbands in a $0.41\degree$ TBG at high magnetic flux. The magnetic-field-induced strong Coulomb interactions, ideal isolation, and flattening of Hofstadter subbands produce strong electronic correlation to induce various isospin-symmetry-broken correlated insulating states at partial fillings of moiré unit cell. Our results establish Hofstadter subbands in minimally twisted TBG as a platform for studying electronic correlations that open new opportunities as compared to the magic-angle TBG.

\begin{figure*}
  \centering
  \includegraphics[width= 1\textwidth]{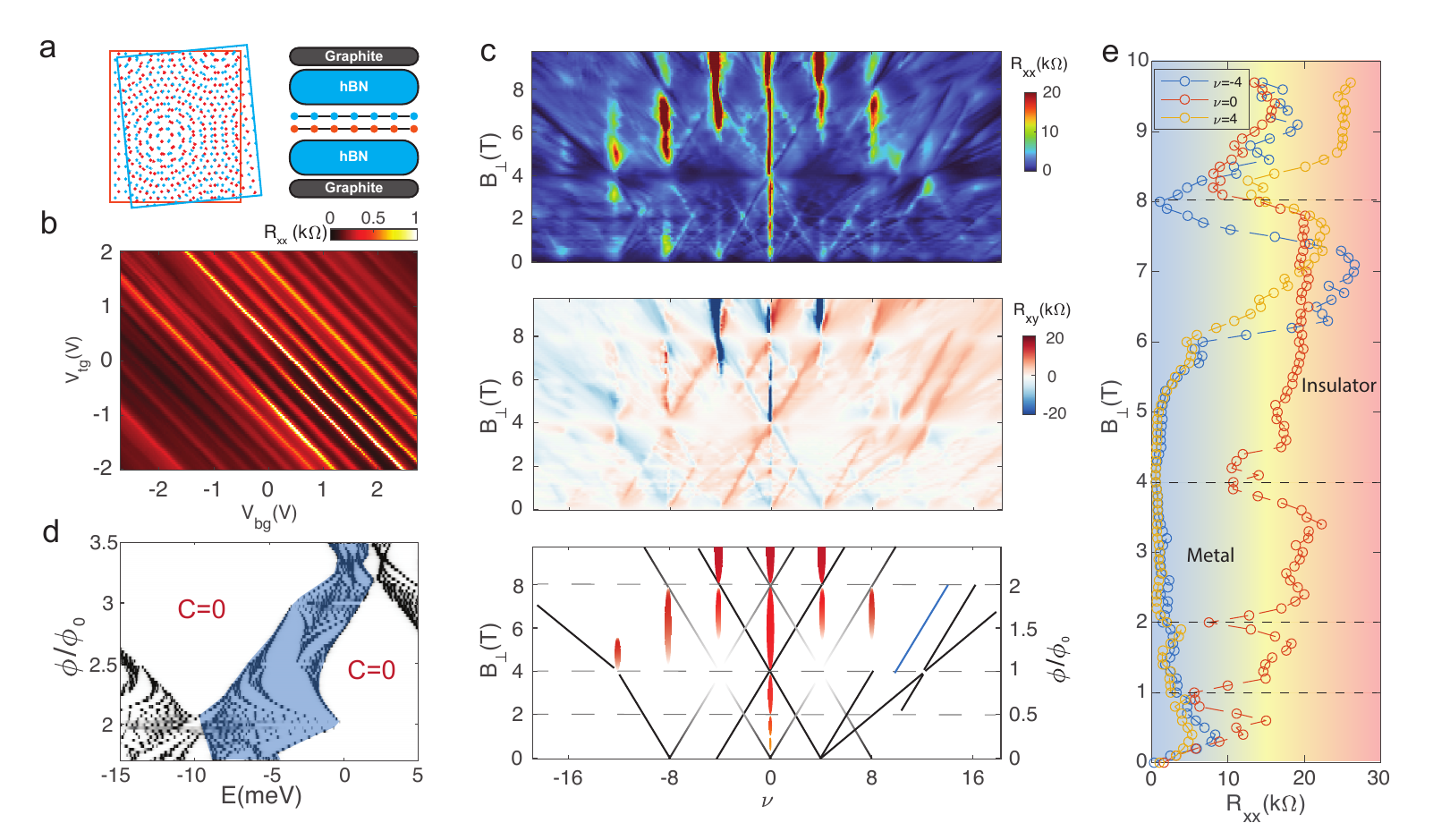}
  \caption{\textbf{Hofstadter butterfly and band gap in small-angle twisted bilayer graphene.} \textbf{a}, Schematics of moiré pattern and device structure. \textbf{b}, Dual gate dependence of four-probe resistance. \textbf{c}, Quantum oscillations in magnetic field characterized by longitudinal magneto resistance $R_{xx}$ (top panel) and Hall resistance $R_{xy}$ (middle panel). The bottom panel illustrates band insulators (red) and orbital Landau levels (black). The right y-axis shows the corresponding magnetic flux. \textbf{d}, Non-interacting numeric simulation of Hofstadter subband structure for $0.41\degree$ TBG. Single-particle band gaps are illustrated with the Chern number $C=0$. The Hofstadter subbands are marked by blue shading. \textbf{e}, Metal-insulator phase transition mediated by the perpendicular magnetic field. The lincuts are acquired from (\textbf{d}) (top panel) at moiré unit cell filling $\nu$=-4 (blue), $\nu$=0 (red), and $\nu$=4 (yellow).}
\label{fig1}
\end{figure*}

\begin{figure*}
  \centering
  \includegraphics[width= 1\textwidth]{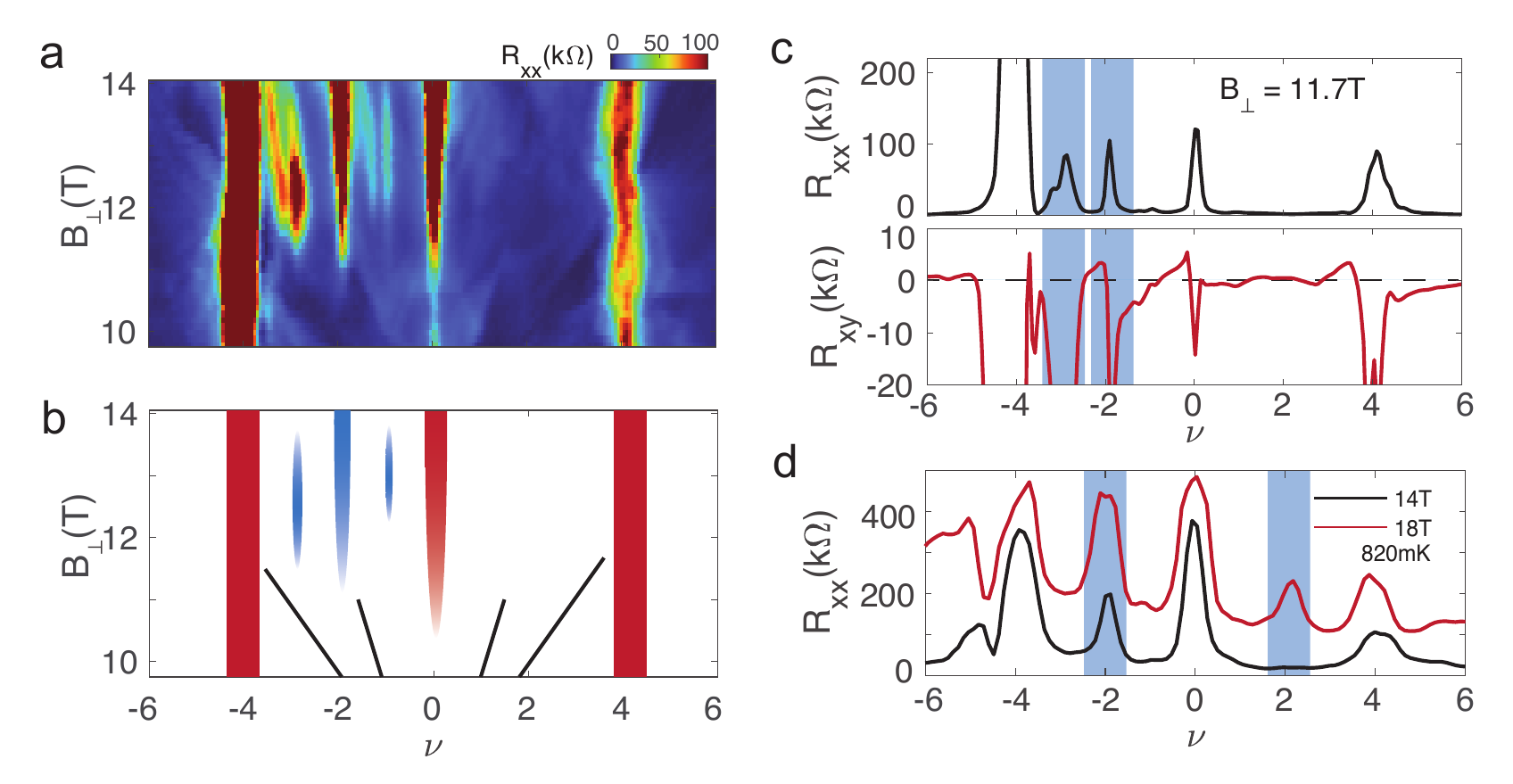}
  \caption{\textbf{Correlated insulators and phase transition in Hofstadter subbands.} \textbf{a}, Longitudinal magneto resistance $R_{xx}$ as a function of charge filling $\nu$ and magnetic field $B_\perp$. \textbf{b}, Schematic illustration of states in (\textbf{a}). Correlated insulators at $\nu$=-3, -2 and -1, band insulators at $\nu$=-4, 0 and 4, and orbital Landau levels are represented by blue, red, and black shadings, respectively. \textbf{c}, Linecuts of $R_{xx}$ and $R_{xy}$ as a function of charge filling $\nu$. \textbf{d}, Emergent half-filling correlated insulator on electron side in a high magnetic field.}
\label{fig2}
\end{figure*} 

\section*{Gapped Hofstadter subbands}

We use a standard dry van der Waals transfer technique to fabricate our stacking, which is composed of dual graphite gates, twisted graphene bilayers, and hexagonal boron nitride dielectric layers. We produced our TBG stackings with a targeting twist angle less than $1\degree$, which is prone to reduction during the assembling process. In this study, we focus on TBG with $0.41\degree$ twist angle, which is close to the predicted ‘second magic angle’ (SMA) (depending on the parameters of lattice reconstruction and interlayer hopping, SMA vary in the interval of $0.2\degree - 0.5\degree$)\cite{Song2019AllTopological, Tarnopolsky2019OriginGraphene}. We carried out four-terminal conductance measurements at a fridge temperature of around 8mK unless specified. SMA is speculated to host electronic correlation physics resembling that at the $1.1\degree$ magic angle. However, our results show quite distinct behaviors at these two magic angles. At $1.1\degree$ twist angle, TBG exhibits single-particle band gaps at one full filling of the moiré unit cell (corresponding to four electrons or holes in one moiré unit cell) and thus flat bands isolated from remote dispersive bands. Unlike $1.1\degree$ TBG, our $0.41\degree$ system features various metallic resistance peaks over a large doping regime in four-terminal transport measurements (Fig.\ref{fig1}b). These peaks correspond to multiple full fillings of the moiré unit cell but without gap openings. Our non-interacting numeric simulations show multiple energy bands, despite a small energy scale of the order of 10 meV, overlapping with each other (Fig.\ref{figS1}). As a result, the effective bandwidth is enlarged to the extent that the correlated insulator phase at partial fillings is no longer favored. The dual gate measurements show a behavior independent of the displacement field, consistent with the strong interlayer coupling in TBG with a small twist angle(Fig.\ref{fig1}b).

Our device exhibits a clear Hofstadter spectrum featuring intercepted cyclotron Landau levels from different fillings at rational magnetic flux  $\Phi=\Phi_0/q$, where $q$ is integers and $\Phi_0=h/e$ is the quantum magnetic flux. The well-determined magnetic flux helps us to reconfirm the twist angle via magnetic field $B_0=4$T for one quantum magnetic flux. Distinct from hetero-bilayer superlattices like graphene/hBN moiré structure, where moiré wavelength is limited within a specific range due to the lattice mismatch, twisted homo-bilayer can host extremely large moiré wavelength at small twist angles. Therefore, twisted homo-bilayer has the advantage of reaching high magnetic flux per moiré unit cell with lower magnetic fields that are easily available in the laboratory. Fig.\ref{fig1}c show clear signs of two flux quanta where Landau levels prominently cross with each other. The resetting of Landau levels at rational magnetic quantum flux, shown by approximately zero Hall resistance $R_{xy}$, originates from a renormalized zero effective magnetic field.  

Strikingly, metallic states of various full fillings at zero field develop into insulating states at the finite magnetic field, as characterized by quite high longitudinal resistance $R_{xx}$ and sign change of Hall resistance $R_{xy}$ over doping (Fig.\ref{fig1}d, see also Fig.\ref{figS3}). Respectively, charge neutrality point (CNP) shows immediate gap opening in the field, while full fillings of $\nu=n/n_0=±4,±8$ require magnetic flux $\Phi=\Phi_0$ (here $n$ is charge density and $n_0$ indicates charge density when moiré unit cell is filled with one electron). In Fig.\ref{fig1}d, our non-interacting calculations of the Hofstadter spectrum through the Lanczos recursion method \cite{Wu2021LandauSuperlattices, Guan2022ReentrantFluxes, Wu2018WannierTools:Materials} predicts such single-particle band gaps in the magnetic field with Chern number $C=0$ (see more calculations in Fig.\ref{figS2}), however, with a contrasting behavior of nearly closing when $\Phi>3\Phi_0$ at $\nu=0$ compared to experimental results. The low-energy band, considered roughly as renormalized mini Dirac cones, is associated with orbital cyclotron energy, spin/orbital Zeeman splitting, and Coulomb interaction when subjected to the perpendicular magnetic field. Like in graphene, the field-induced gap at CNP is dominated by the Coulomb interaction, featuring continuous increment in a magnetic field and hosting complicated ground states of charge density wave or intervalley coherence\cite{Young2012SpinIngraphene, Liu2022VisualizingFerromagnet}. This may interpret the inconsistency of field-dependence of the CNP gap between our theoretical calculation and experimental results.

We also note these gapped states are modulated by the crossing of the electron and hole cyclotron Landau levels at rational flux, compatible with our zero-field overlapped band structure and thus implying a candidate ground state of excitonic insulator at finite magnetic field. We note previous studies proposed a fragile topology derived from identical helicity of Dirac bands in the moiré Brillouin zone to interpret Landau level crossing with band gaps at CNP and full fillings \cite{Lian2020LandauTopology, Lu2021MultipleAngle}. The exact mechanism underlying this crossing is beyond the scope of this work.

\begin{figure}
  \centering
  \includegraphics[width= 0.5\textwidth]{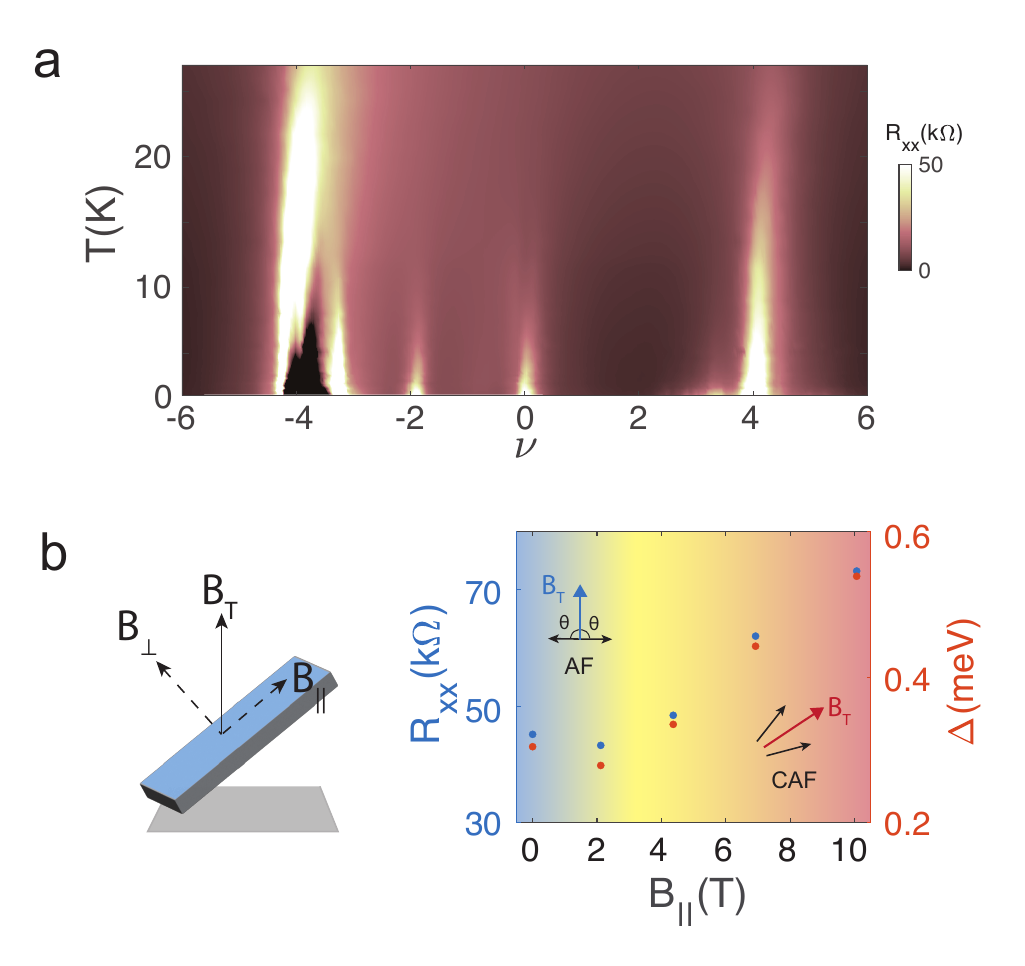}
  \caption{\textbf{Spin texture of correlated insulator at $\nu=2$.} \textbf{a}, Temperature dependence of longitudinal resistance $R_{xx}$ as a function of charge filling $\nu$ at perpendicular magnetic field $B_{\perp}=11.7$T. \textbf{b}, In-plane magnetic field dependence of correlated insulator resistance at 1.5K and thermal-activation gap $\Delta$ at $\nu=2$. The left panel shows a rotated sample measurement scheme. The total magnetic field $B_T$ is projected to in-plane $B_{\parallel}$ and out-of-plane $B_{\perp}$ component. $B_{\perp}$ is fixed at 12T. The inset figures show the spin texture transition from an antiferromagnetic to a canted antiferromagnetic state.}
\label{fig3}
\end{figure}

\begin{figure*}
  \centering
  \includegraphics[width= 0.7\textwidth]{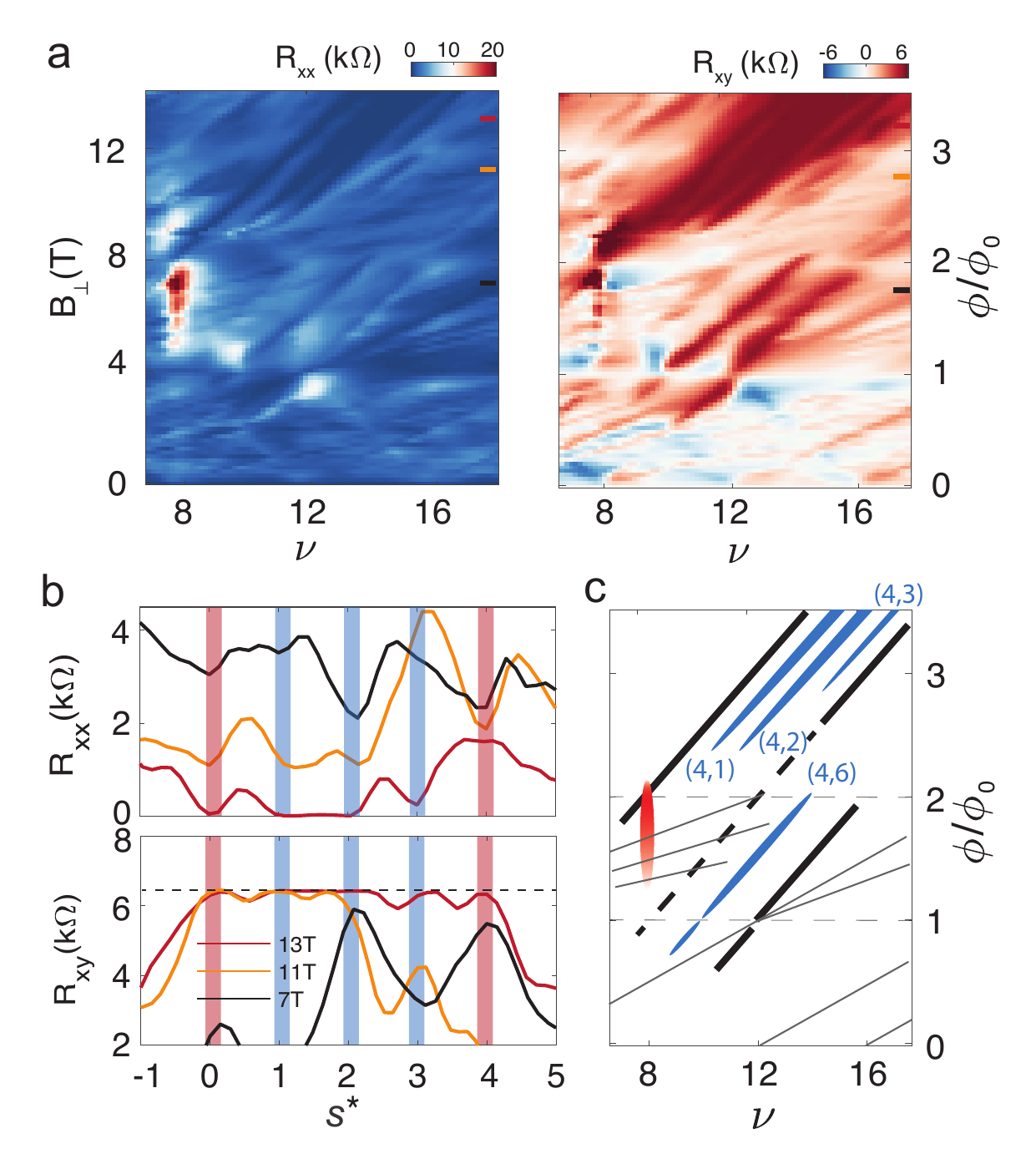}
  \caption{\textbf{Isospin-symmetry-broken correlated insulators with nonzero Chern number.} \textbf{a}, Mapping plot of longitudinal magnetoresistance $R_{xx}$ and Hall resistance $R_{xy}$ as a function of perpendicular magnetic field $B_\perp$ and charge filling $\nu$. \textbf{b}, Linecuts of $R_{xx}$ and $R_{xy}$ acquired from (\textbf{a}) with respect to renormalized charge filling $s^{\star}$. Here $s^{\star}=s$ and $s^{\star}=s-4$ for $B_\perp=13,11$T and $B_\perp=7$T, respectively. $s$ follows the Diophantine equation $\nu=C\Phi/\Phi_0+s$. The $R_{xx}$ linecuts are consecutively shifted upward by 1k$\ohm$. The dashed line in the bottom figure represents the Hall resistance of $4h/e^2$. \textbf{c}, Schematic illustration of states in (\textbf{a}). Band gaps with Chern number $C=0$, $C=4$, and correlated gaps are illustrated by red, black, and blue shadings, respectively.}. 
\label{fig4}
\end{figure*}

\section*{Correlated insulators in Hofstadter subbands}

Upon approaching $\Phi=3\Phi_0$ we found that the quantum oscillations corresponding to orbital Landau levels embedded between band insulators at $\nu=0, \pm4$ disappear. When the moiré unit cell is partially filled with holes, these quantum oscillations give way to resistive states, which show no slopes of $dn/dB_\perp$ in the fan diagram (Fig.\ref{fig2}a). Specifically, we observe these resistive states at half filling ($\nu=-2$, two holes per moiré unit cell) and quarter fillings ($\nu=-1, -3$, one and three holes per moiré unit cell, respectively). These states emerge with a hierarchy of required magnetic fields. Resistive states also appear at fractional but non-fixed $\nu$, which might be associated with changed singularities (like van Hove singularities) in Hofstadter subbands modulated by magnetic field. On the electron side, no state is found at partial fillings below $B_\perp=14$T, indicating a considerable particle-hole symmetry breaking in the electronic band structure, which is quite common in twisted and non-twisted crystalline multilayer graphene \cite{Chen2019EvidenceSuperlattice, Shen2020CorrelatedGraphene, Liu2020TunableGraphene, Cao2020TunableGraphene, Chen2021ElectricallyGraphene}. Increasing the field to $\Phi>4\Phi_0$, we find a half-filling state at the electron side eventually appears  (Fig.\ref{fig2}d) (the high-resistance background is due to the resistive state of $\nu=0$ at high magnetic field which yields a bad contact).

To further elucidate these states at partial fillings, the Hall resistance $R_{xy}$ exhibiting sign change at integer filling $\nu=2$ as shown in Fig.\ref{fig2}c, points to Fermi surface topology transition with gap opening. We note the Hall resistance $R_{xy}$ is obviously shifted by colossal longitudinal $R_{xx}$ component since the misalignment of opposite Hall bars is inevitable. Such an insulating state is further evidenced by the electronic thermal activation behavior of increasing resistance with decreasing temperature at $\nu=2$ (Fig.\ref{fig3}a).

The states at partial fillings $\nu=-1,\pm2,-3$ of Hofstadter subbands unambiguously indicate SU(4) spin-valley symmetry breaking induced by strong electronic correlation. To understand the underlying mechanism for strong electronic correlation in Hofstadter subbands, we consider the typical Stoner ferromagnetism model involving competing Coulomb exchange interaction $U$ and density of state at Fermi level $D(E_F)$. In the Stoner picture, when $D(E_F)\cdot U>1$ is satisfied, the isospin flavor symmetry will be spontaneously broken, forming spin or valley-polarized correlated states. The magnetic field leads to high $U$ and $D(E_F)$ to satisfy Stoner criteria, as we will address in the following.

\begin{figure*}
\renewcommand{\thefigure}{S1}
  \centering
  \includegraphics[width= 0.9\textwidth]{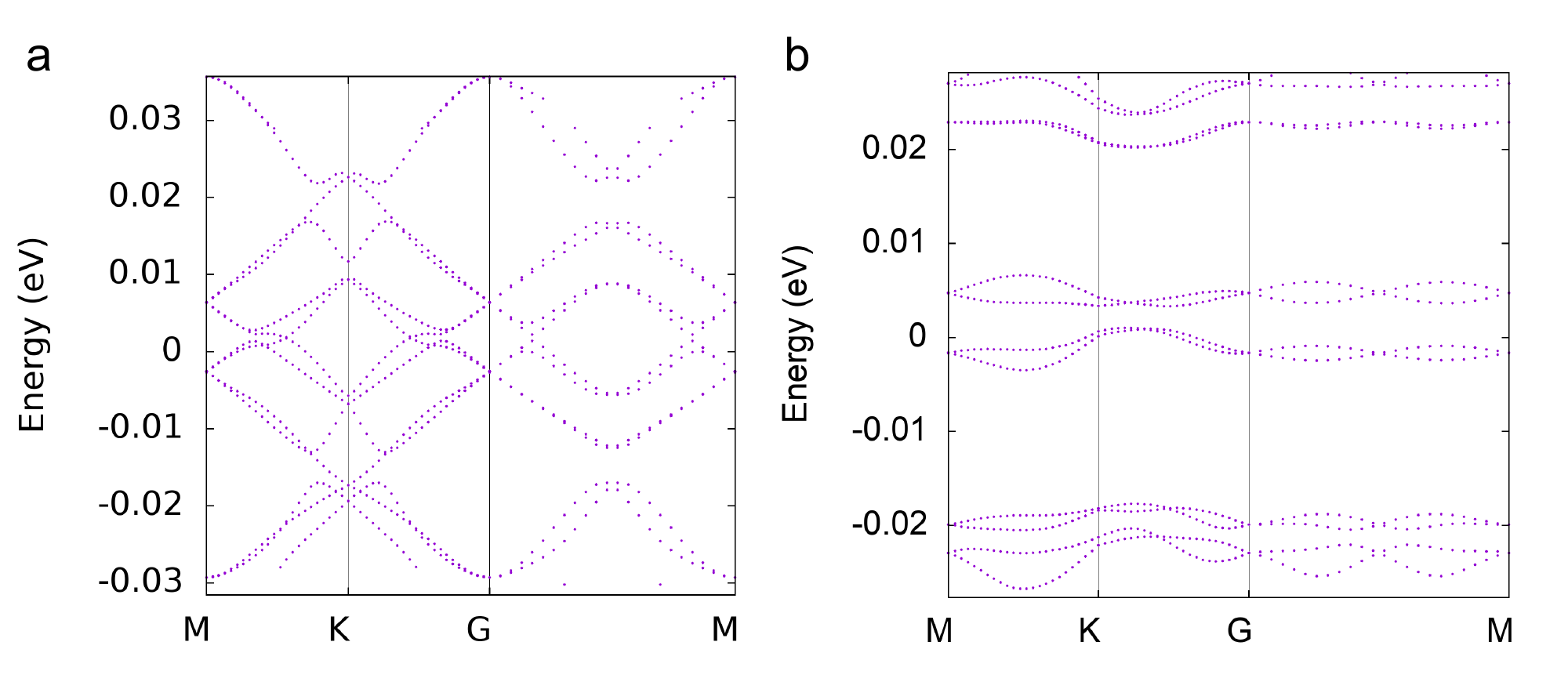}
  \caption{\textbf{Non-interacting band structures of TBG in the moiré Brillouin zone at 0.41\degree twist angle.} \textbf{a}, Band structure at $\Phi=0$. \textbf{b}, Band structure at $\Phi=3\Phi_0$. }. 
\label{figS1}
\end{figure*}

An out-of-plane magnetic field $B_\perp$ substantially reconstructs the zero-field low-energy moiré bands. In our case, band reconstruction includes: (1) Hofstadter subbands isolation by gap openings at $\nu=0,\pm4$; (2) reduced bandwidth of Hofstadter subbands. The crucial role of Hofstadter subband isolation for correlated insulators is reflected by the hierarchy that partial fillings $\nu=-1,\pm2,-3$ require preformed gap states at $\nu=0,\pm4$. The bandwidth of Hofstadter subbands is directly related to orbital Landau level gaps inside itself. In Fig.\ref{fig2}a, the disappearance of quantum oscillation (illustrated as black lines in Fig.\ref{fig2}b) in high field signifies reduced orbital Landau level gaps, as Landau level broadening is exclusively induced by the fixed disorder and temperature effect here. Thus, this phase transition directly points to a related flattening of Hofstadter subbands where high $D(E_F)$ is available. In Fig.\ref{figS1}b, the calculated Hofstadter subbands at $3\Phi_0$ show a narrow bandwidth smaller than 10meV. We roughly estimate the on-site Coulomb interaction energy in 0.41$\degree$ TBG at the zero-field limit compared to magic angle \cite{Wong2020CascadeGraphene} by only considering the change of moiré wavelength. It yields on-site Coulomb interaction energy of the same order as the calculated bandwidths, implying a correlated regime for Hofstadter electrons. Despite the real bandwidth or density of the state is hard to precisely acquire theoretically (due to the strong lattice relaxation effect for small angle TBG), we also emphasize the impact of Coulomb exchange interaction $U$ between different isospins. Since $U$ is proportional to $e^2/l_B$ where $l_B=\sqrt{\hbar/eB_\perp}$ is the magnetic length, the Stoner criteria can also be satisfied with a much enhanced $U$ in a high magnetic field. Observing correlated insulators of partial fillings in higher magnetic flux is consistent with the above analysis. In contrast, in magic-angle TBG, isospin symmetry breaking for Hofstadter subbands is realized at a low magnetic field due to inherited strong electronic correlation intrinsic to zero fields \cite{Young2012SpinIngraphene}. Hofstadter subbands flattening and enhanced Coulomb exchange interaction by magnetic flux, therefore providing an avenue for realizing strong correlations in a system that doesn’t inherently host such a scenario.

\section*{Spin flavor ordering at half filling of Hofstadter subbands}

The isospin symmetry breaking leads to intricate isospin flavor ordering. Subjected to the competing spin Zeeman effect and Coulomb exchange interaction in a perpendicular magnetic field, the half-filling correlated insulator can be either a spin-unpolarized state due to an antiferromagnetic exchange interaction or a spin-polarized state due to the spin alignment by a magnetic field. We apply a tilted magnetic field $B_T$, whose out-of-plane projection $B_{\perp}$ is fixed and in-plane $B_{\parallel}$ component is varied. As the in-plane magnetic field can only alter the electron's spin, not its orbit, the response to the in-plane component of the tilted magnetic field can thus help to unveil the spin flavor ordering of the ground state. Fig.\ref{fig3}b shows the four-probe resistance and thermal-activation gap (fitted with Arrhenius law) at half filling. They exhibit non-monotonic behavior with respect to the in-plane field, showing a minimum at $B_{\parallel}$ around 2 T to 4 T. The decreasing of the correlated gap by a small in-plane field reveals an in-plane spin-unpolarized ground state with antiferromagnetic (AF) ordering at $B_{\perp}=B_T$. The larger in-plane field alters the spin inclined to in-plane ferromagnetic ordering, forming a canted antiferromagnetic (CAF) state with respect to the direction of the tilted field (as shown in Fig.\ref{fig3}b). The canted angle $\theta$ is determined by the ratio of Zeeman energy $g\mu_BB_T$ and the antiferromagnetic exchange interaction, which depends only on $B_\perp$ (here, $g$ is bare gyromagnetic ratio, $\mu_B$ is Bohr magneton).

\section*{Correlated gaps with non-zero Chern number}

Spanning from $\nu=0$ to $\nu=-4$, all the insulator states exhibit zero Chern number as a characteristic of zero $dn/dB$ in the phase diagram shown in Fig.\ref{fig2}a according to the Streda formula $C=\frac{h}{e}\cdot\frac{dn}{dB}$. In contrast to a common expectation that magnetic field endows Hofstadter subband with nonzero Chern numbers\cite{Saito2021HofstadterGraphene, Spanton2018ObservationHeterostructure}, here the Hofstadter subband, as embedded between gaps with zero Chern numbers at $\nu=0$ and $\nu=-4$, has a net Chern number $\Delta C=C(\nu=0)-C(\nu=-4)=0$. For correlated gap at partial fillings $\nu$, its Chern number is  $C(\nu)=C(\nu=-4)+\nu\Delta C/4$, which turns out to be always zero. In Fig.\ref{fig4}, we depict symmetry-broken correlated states that correspond to filling the moiré unit cell with one, two, or three electrons in a regime of magnetic flux $\Phi>2\Phi_0$ and two electrons for $\Phi<2\Phi_0$. These states show more pronounced symmetry breaking in higher magnetic flux, which is compatible with the above picture of band flattening and enhanced Coulomb interaction by a magnetic field. However, they all have a Chern number $C=4$ with edge states as characterized by the zero longitudinal magnetoresistance $R_{xx}$ and Hall resistance $R_{xy}$ plateau of approximately $4h/e^2$ (Fig.\ref{fig4}b). Described by Diophantine equation $\nu=C\Phi/\Phi_0+s$, these states are characterized by $(C,s)=(4,1),(4,2) (4,3)$ and $(4,6)$, respectively. Having the same origin as correlated insulators in the first valence Hofstadter subbands, these correlated states appear by splitting higher-energy flat conduction Hofstadter subbands via the isospin symmetry breaking. The net Chern number of these Hofstadter subbands is $\Delta C=0$ as well, as determined by the same Chern number $C=4$ for the two adjacent orbital Landau level gaps in the Hofstadter spectrum. Splitting these Hofstadter subbands yields correlated gaps with Chern number $C=4$. 

To conclude, the flat Hofstadter subbands in small-angle twisted bilayer graphene that are universal in various magnetic flux and charge fillings provide a new pathway for investigating strongly correlated electronic phases. The phase transition associated with isospin symmetry breaking observed in our work lays the foundation to study correlated Hofstadter spectrum and achieve intriguing fractional Chern insulating states in Hofstadter subbands with nonzero net Chern number$\Delta C\neq0$ if possible \cite{Xie2021FractionalGraphene, Spanton2018ObservationHeterostructure}. The intricate ground states and isospin flavor ordering for correlated insulators at partial fillings and even band insulators in Hofstadter subbands appeal to more theoretical efforts. Further experimental investigations through applying higher magnetic flux, local compressibility measurements\cite{Xie2021FractionalGraphene}, etc., will help to shed light on these exotic quantum phases.

\bibliography{references_1,references_2}

\begin{acknowledgments}
We thank Ady Stern for the fruitful discussions. C.S. acknowledges the cleanroom facilities at Cmi (EPFL). We acknowledge the support of the HFML, member of the European Magnetic Field Laboratory (EMFL). M.B. acknowledges the support of SNSF Eccellenza grant No. PCEGP2\underline{~}194528, and support from the QuantERA \RomanNumeralCaps{2} Programme that has received funding from the European Union’s Horizon 2020 research and innovation program under Grant Agreement No 101017733. Y.G. and O.V.Y. acknowledge support from the Swiss National Science Foundation (grant No. 204254). Computations were performed at the Swiss National Supercomputing Centre (CSCS) under project No. s1146 and the facilities of the Scientific IT and Application Support Center of EPFL. K.W. and T.T. acknowledge support from the JSPS KAKENHI (Grant Numbers 20H00354 and 23H02052) and World Premier International Research Center Initiative (WPI), MEXT, Japan.
\end{acknowledgments}

\textbf{\begin{center}Author Contributions\end{center}}

C.S. fabricated devices and performed measurements (\textless14T) and data analysis. Y.G. and O.V.Y. carried out numeric simulations. C.S.performed the high-field measurements with the help from D.P., Z.Z., P.B. and S.W.. T.T. and K.W. provided hBN crystals. M.B. supervised this project. C.S. wrote the paper with input from Y.G., O.V.Y. and M.B.. 

\textbf{\begin{center}Competing interests\end{center}}
The authors declare no competing interests.

\textbf{\begin{center}Data availability\end{center}}
The data supporting the findings of this study are available from the corresponding author upon reasonable request.

\newpage
\clearpage
\onecolumngrid

\section*{Supplementary Information}
\subsection{Device fabrication and transport measurement}
We made stackings using the van der Waals assembly technique. Graphite, hexagonal boron nitride, and graphene flakes were exfoliated onto SiO$_2$(285nm)/Si++ substrate and picked up at 100$^\circ$C with a propylene carbonate (PC) film which is placed on a polydimethyl siloxane (PDMS) stamp. We twisted graphene bilayers with a targeting twist angle less than 1$^\circ$. Stackings are fabricated with a Hall bar configuration via e-beam lithography and SF$_6$/O$_2$ plasma etching and contacted with Cr/Au metal. Measurements are done with lock-in amplifiers in Bluefors dilution fridge at a base temperature of 8mK unless specified.

The twist angle is identified through the magnetic flux quanta $\Phi_0$. $\Phi_0$ and the corresponding magnetic field $B_0$ are acquired from the Hofstadter spectrum. The twist angle is therefore obtained via $\theta=\sqrt{\sqrt{3}B_0a^2/(2\Phi_0)}$ (here $a=0.246nm$ is graphene lattice constant).

\subsection{Band-structure calculations}

\subsubsection{Tight-binding Hamiltonian} 
The electronic structure calculations are performed with the tight-binding Hamiltonian accounting for the atomistic details of twisted bilayer graphene in free-electron formalism,
\begin{equation*}
H=\sum_{i.,j}t_{ij}c^\dagger_ic_j
\end{equation*}
where $c_i$($c^{\dagger}_i$) are the annihilation (creation) operators of the $p_z$-orbital electron states on the $i$-th carbon atom. The hopping integral $t_{ij}$ between two $p_z$ orbitals is expressed as a sum of $\sigma$-type and $\pi$-type couplings following the Slater-Koster formalism \cite{Gargiulo2018StructuralGraphene, Zhu2012StructureWrinkles,Slater1954SimplifiedProblem}
\begin{equation*}
t_{ij}=V^\pi_0 \exp(-\frac{r-a_0}{r_0}) \sin^2\theta+V^\sigma_0exp(-\frac{r-d_0}{r_0}) \cos^2\theta,
\end{equation*}
where $r=|\textbf{r}_{ij}|$ and $\theta$ is the angle between $\textbf{r}_{ij}$ and $\textbf{e}_z$, $\theta=\angle(\textbf{r}_{ij},(0,0,1))$. The numerical parameters have been adapted from Ref.\cite{nam2017lattice}; hence $V^\pi_0=-2.7$~eV is the hopping integral between nearest-neighbor (NN) atoms in the same layer; $V^\sigma_0=0.48$~eV is the hopping integral between the aligned atoms in different layers. Here, $a_0=0.142$~nm and $d_0=0.334$~nm correspond to the nearest neighbor distance and the interlayer distance; $r_0=0.148a$ (we use $a=\sqrt{3}a_0$ as the lattice constant of graphene) is the hopping strength decay length. The cutoff in real space is set to 9.84$\AA$, outside which the coupling vanishes.
It is known that the relaxation effect has a significant effect on the band structures of TBG \cite{nam2017lattice,nakatsuji2023multiscale}.
We use the classical force fields package LAMMPS \cite{LammpsJCP} to relax the atomic structure.
The classical force field contains the bond-order potential\ \cite{SRpotentional} as well as the modified version of the Kolmogorov–Crespi registry-dependent potential for describing the interlayer interactions\ \cite{KolmogorovPRB}.
The Kolmogorov–Crespi long range potential is fitted to DFT results\ \cite{Gargiulo2018StructuralGraphene}.

\subsubsection{Chern numbers in Hofstadter gaps}  
\begin{figure*}
\renewcommand{\thefigure}{S2}
  \centering
  \includegraphics[width= 0.5\textwidth]{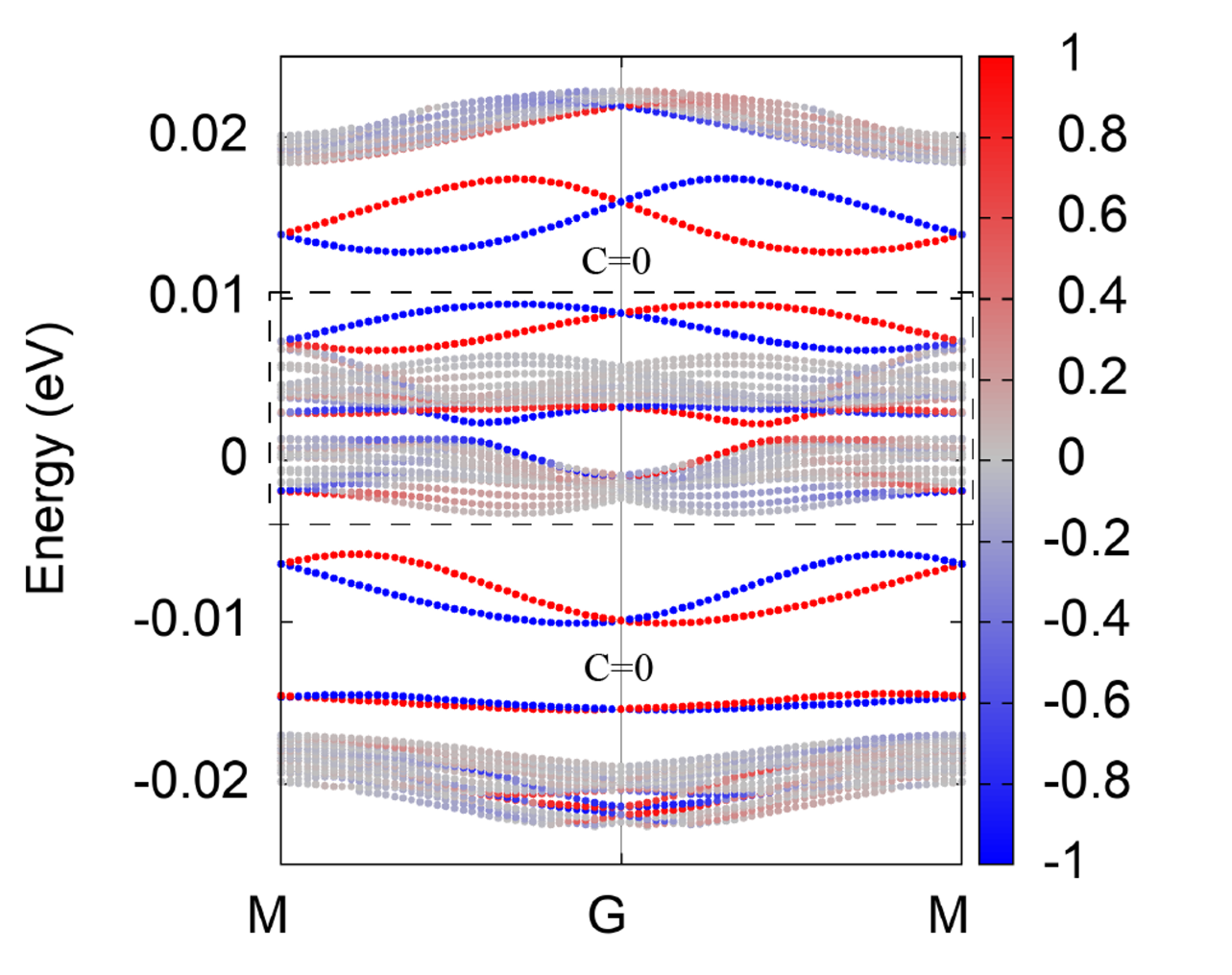}
  \caption{\textbf{Example of the edge state band structure plot at $\Phi=3\Phi_0$}. The color of the points encodes the localization of the states on one of the edges. The dashed box highlights flat Hofstadter subbands.}. 
\label{figS2}
\end{figure*}

\begin{figure*}
\renewcommand{\thefigure}{S3}
  \centering
  \includegraphics[width= 0.7\textwidth]{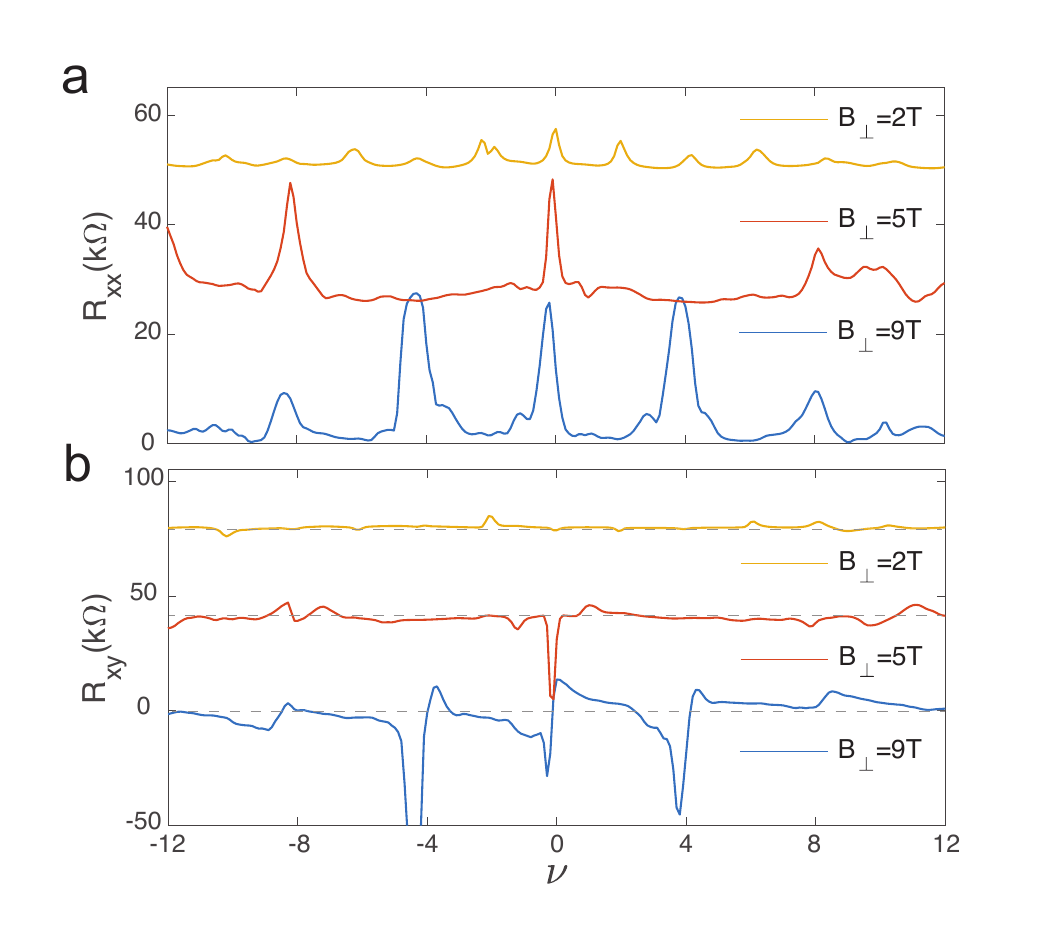}
  \caption{\textbf{Extended data for band gaps}. Longitudinal magneto-resistance $R_{xx}$ (\textbf{a}) and Hall resistance $R_{xy}$ (\textbf{b}) as a function of charge filling $\nu$ at $B_\perp=$2T, 5T and 9T. $R_{xx}$ and $R_{xy}$ for different perpendicular magnetic fields are sequentially shifted by 25k$\ohm$ and 40k$\ohm$, respectively.}. 
\label{figS3}
\end{figure*}

The magnetotransport properties of the system are obtained by Hofstadter spectrum calculations through the Lanczos recursion methods. We calculate the chiral edge states in the Hofstadter gaps to probe the quantized Hall conductance in the topological gaps. Defining the external magnetic field $\textbf{B}=B\textbf{e}_z$, finite width is taken along the $y$-direction, while the $x$-direction is periodic. For example, in Figure~\ref{figS2} we present the band structure plot in such a configuration for $\Phi=3\Phi_0$. The localization of the eigenstates on the edge is represented with the color-coding. The number of edge states on each edge corresponds to the Chern number in the gap.

\subsection{Twist-angle homogeneity}
To identify the twist-angle homogeneity, we measure different pairs of electrodes. Fig.S4 shows the four-probe resistance. It's obvious that the charge density corresponding to full fillings of moiré unit cell doesn't vary much among different regions, indicating high twist-angle homogeneity with variation less than 0.007 $\degree$.

\subsection{More data for correlated insulators}
Fig.S5 shows extended data acquired from the device with twist angle $\theta=0.37\degree$ (we examine the twist angle via the magnetic flux quanta at 3.6T). All the features including the band gaps and Hofstadter butterfly resemble to the device of $\theta=0.41\degree$ discussed in the main text. We observed the resistive state at $\nu=2$ when $B_\perp>10$T. 

\newpage
\clearpage
\onecolumngrid

\begin{figure*}
\renewcommand{\thefigure}{S4}
  \centering
  \includegraphics[width= 0.9\textwidth]{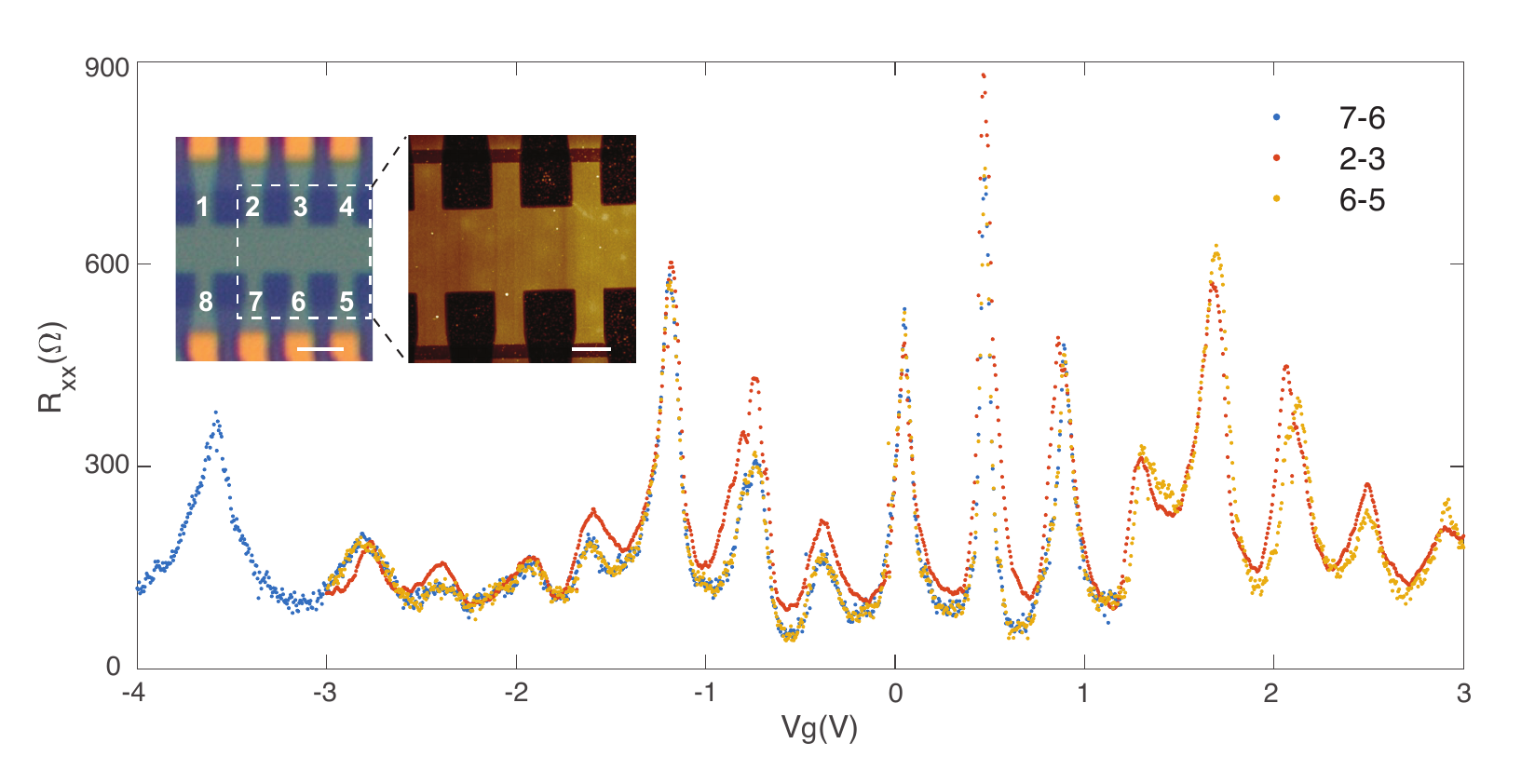}
  \caption{\textbf{Twist angle homogeneity}. Four-probe resistance $R_{xx}$ data acquired from different pairs of electrodes at zero field. The inset figures show the optical and atomic force microscopic images of the device.The error bars for two figures are 2$\mu m$ (left) and 1$\mu m$ (right).}. 
\label{figS4}
\end{figure*}

\begin{figure*}
\renewcommand{\thefigure}{S5}
  \centering
  \includegraphics[width= \textwidth]{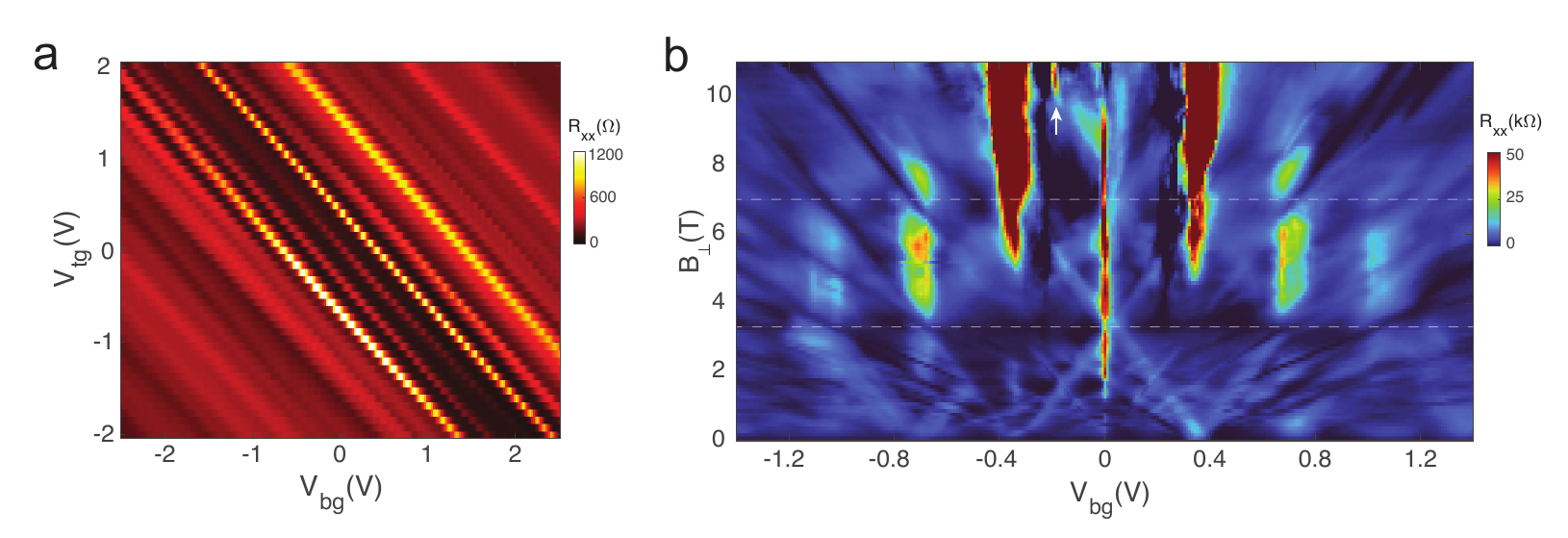}
  \caption{\textbf{Extended data acquired from another device of $\theta=0.37\degree$}. \textbf{a}, Dual-gate mapping of four-probe resistance $R_{xx}$. \textbf{b}, Longitudinal magneto-resistance as a function of back-gate voltage and perpendicular magnetic field. Top-gate voltage is fixed at 0V. The dash lines mark the magnetic flux quanta $\Phi=\Phi_0$ at 3.4T and $\Phi=2\Phi_0$. The arrow points to the emergent half-filling correlated insulator. Both figures are obtained at 240mK.}. 
\label{figS5}
\end{figure*}

\begin{figure*}
\renewcommand{\thefigure}{S6}
  \centering
  \includegraphics[width= 0.6\textwidth]{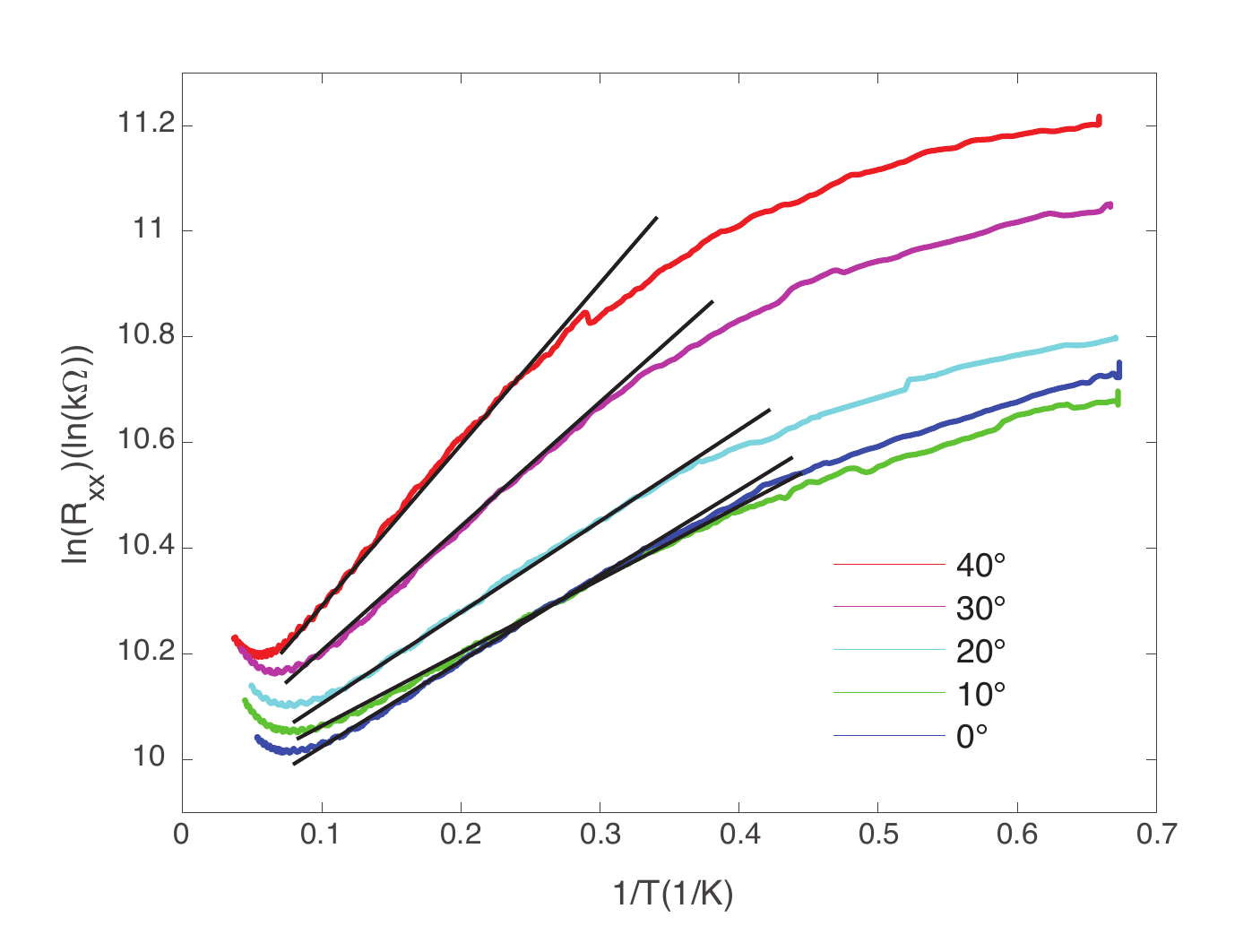}
  \caption{\textbf{Arrhenius fitting of thermal activation gap at $\nu=2$}. The original temperature dependence of four-probe resistance is depicted in the form of logarithm as a function of $1/T$. The device was rotated by 0\degree, 10\degree, 20\degree, 30\degree and 40\degree, corresponding to $B_\parallel=$0T, 2.12T, 4.37T, 6.93T and 10.07T, respectively when $B_\perp$ is fixed at 12T. }. 
\label{figS6}
\end{figure*}

\begin{figure*}
\renewcommand{\thefigure}{S7}
  \centering
  \includegraphics[width= 0.6\textwidth]{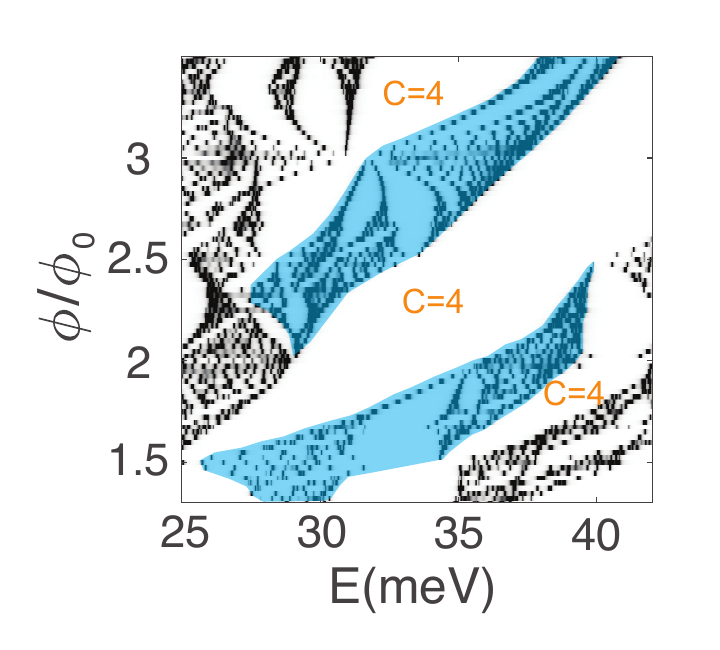}
  \caption{\textbf{Non-interacting tight binding calculation of higher-energy Hofsadter subbands}. The Hofstadter subbands are depicted by blue shadings, adjacent to the $C=4$ Chern gaps.}. 
\label{figS7}
\end{figure*}

\end{document}